\documentclass{article}
\usepackage{graphicx} 

\title{A verified implementation of the Misra and Gries edge coloring algorithm}
\author{Arohee Bhoja}
\date{2025}

\usepackage{amsmath}
\usepackage{enumitem}
\usepackage[sorting=none, backend=biber]{biblatex}
\setlist[itemize,1]{label=-} 
\setlist{nolistsep, itemsep=4pt}
\usepackage{amsthm}
\usepackage[noend]{algpseudocode}
\usepackage{algorithm}

\usepackage[utf8]{inputenc}
\usepackage{listings}
\usepackage{amssymb}

\usepackage[english]{babel}
\usepackage[autostyle]{csquotes}

\usepackage{color}
\definecolor{keywordcolor}{rgb}{0.7, 0.1, 0.1}   
\definecolor{tacticcolor}{rgb}{0.0, 0.1, 0.6}    
\definecolor{commentcolor}{rgb}{0.4, 0.4, 0.4}   
\definecolor{symbolcolor}{rgb}{0.0, 0.1, 0.6}    
\definecolor{sortcolor}{rgb}{0.1, 0.5, 0.1}      
\definecolor{attributecolor}{rgb}{0.7, 0.1, 0.1} 

\newcommand{\code}[1]{\lstinline{#1}}

\usepackage[many]{tcolorbox}

\newtcolorbox{myquotebox}{
  enhanced,
  blank,
  left skip=8pt,
  borderline west={1pt}{-7pt}{lightgray}}

\algblockdefx[FunDef]{FunDef}{End}[3]
{\textbf{function} \textsc{#1} #2
    \vspace{-4pt}
    \begin{myquotebox}
        #3
    \end{myquotebox}
    \vspace{-4pt}}
[0]{}

\algblockdefx[Fun]{Fun}{End}
[2]{\textbf{function} \textsc{#1} #2 $\defeq$}
[0]{}

\algcblockdefx[Where]{Fun}{Where}{End}
[2]{\textbf{where} \textsc{#1} #2 $\defeq$}
[0]{}

\makeatletter
\ifthenelse{\equal{\ALG@noend}{t}}%
  {\algtext*{End}}
  {}%
\makeatother

\makeatletter
\newcommand*{\defeq}{\mathrel{\rlap{%
                     \raisebox{0.3ex}{$\m@th\cdot$}}%
                     \raisebox{-0.3ex}{$\m@th\cdot$}}%
                     =}
\makeatother

\begin{document}

\maketitle

\begin{abstract}
Vizing's theorem states that every simple undirected graph can be edge-colored using fewer than $\Delta + 1$ colors, where $\Delta$ is the graph’s maximum degree. The original proof was given through a polynomial-time algorithmic procedure that iteratively extends a partial coloring until it becomes complete. In this work, I used the Lean theorem prover to produce a verified implementation of the Misra and Gries edge-coloring algorithm, a modified version of Vizing’s original method (available as source code \cite{mycode}). The focus is on building libraries for relevant mathematical objects and rigorously maintaining required invariants.
\end{abstract}

\section{Introduction}

Let $G$ be a simple undirected graph. We denote the maximum degree of $G$ by $\Delta(G)$, or simply $\Delta$.

\begin{itemize}

\item A \textit{proper edge coloring} of $G$ maps edges in $G$ to colors such that no two incident edges share a color. This edge coloring is \textit{complete} if every edge in the graph is colored.

\item The \textit{edge chromatic number} of $G$, denoted as $\chi'(G)$, is the minimum number of colors necessary to create a proper edge coloring of $G$. 
\end{itemize}

Immediately, we can conclude that $\Delta \le \chi'(G)$. This is because if two edges are incident, they cannot have the same color, and the degree of a vertex is its number of incident edges. 

In 1965, Soviet mathematician Vadim Vizing described a polynomial-time algorithmic procedure that proved $\chi'(G) \le \Delta + 1$. \cite{vizing} It turns out that this is the tightest bound we can achieve with a deterministic polynomial-time algorithm, since it is NP-complete to determine, for a given graph $G$, whether $\chi'(G) = \Delta$ or $\chi'(G) = \Delta + 1$. \cite{npcomplete}

During the latter half of the 20th century, there were several efforts made to simplify Vizing's original algorithm and its proof of correctness. Bollobás presented a shorter proof in 1979, which was really more of a sketch as it omitted several nontrivial pieces of the argument. \cite{bollobas} In the early 1990s, simplified algorithms were designed by Rao and Dijkstra \cite{raodijkstra} and by Misra and Gries \cite{misragries}, that introduced nomenclature for auxiliary data structures and reduced branching wherever possible. 

However, even after these improvements, the proofs of correctness are necessarily rather complex. There are several auxiliary data structures used in the description of each algorithm, which interact with one another in subtle ways. They must all be shown to maintain certain invariants over several subroutines that are themselves nontrivial. In other words: the algorithm itself does not straightforwardly motivate its proof of correctness, which makes it an interesting case study for formal verification. 

A number of graph theoretic algorithms have been formally verified using both interactive and automated theorem provers. In 2024, Tobias Nipkow verified the Gale–Shapley stable matching algorithm in Isabelle by progressing from abstract specifications to concrete implementations via stepwise refinement. \cite{nipkow} However, this approach does not readily extend to the Misra and Gries algorithm, as its complexity and reliance on auxiliary data structures makes stepwise concretization impractical. Additionally, in 2018, Tarjan’s algorithm for computing the strongly connected components of a directed graph was verified in the deductive program verifier Why3 and the interactive theorem provers Rocq and Isabelle. \cite{tarjan} While valid, these verifications relied on abstract library data types and therefore did not constitute rigorous proofs of computationally efficient implementations.

Earlier work by Joachim Breitner in 2011 implemented the Misra and Gries edge-coloring algorithm in the modeling formalism Event-B and verified it using the Rodin tool. \cite{breitner} Rodin automatically extracts proof obligations from the implementation to be discharged manually in Isabelle. While the resulting formal proof of correctness was sound, it had several limitations: it relied on the assumed soundness of Rodin, and the automatically generated proofs were described as \enquote{write-once, read-never.} Through the use of an interactive theorem prover, the present work seeks to mitigate these issues by establishing a closer correspondence between informal and formal proofs.

\section{The algorithm and its proof of correctness}

We now present the algorithm as it is implemented in the formalization.

Let $V(G)$ denote the vertex set of $G$, and $E(G)$ denote its edge set.

Let $u \in V(G)$, and $c$ be a color. Fix a (partial) proper $(\Delta + 1)$-edge coloring $C$. $c$ is an: 

\begin{itemize}
\item{ \textit{incident color} on $u$ if there exists an edge incident to $u$ that is colored $c$ }
\item{ \textit{free color} on $u$ otherwise. }
\end{itemize}

A \textit{fan} $F = \langle f_1, ... f_k \rangle$ on a vertex $x$ is a nonempty sequence of distinct neighbors of $x$ that satisfies the following properties:
\begin{itemize}
\item{The edge $\{x, f_1\}$ is uncolored.}
\item{\textit{Fan color property:} for every $i < k$, the color of $\{x, f_{i+1}\}$ is free on $f_i$.}
\end{itemize}

When $x$ is clear from context, elements of the fan will be used as shorthand for their corresponding fan edges (e.g. $f_i$ as shorthand for $\{x, f_i\}$).

If we would like to color $\{x, f_1\}$, there is an easy process to do so based on the properties of the fan -- since every fan edge's color is free on its predecessor, we can rotate each color down one edge and leave the last fan edge uncolored.

For every $i$, let $c_i$ denote the color of edge $\{x, f_i\}$. We can now describe this procedure in pseudocode:

\begin{algorithmic}
\FunDef {rotate} {$C[F]$}
    {For every $i < k$, recolor $\{x, f_i\}$ with $c_{i+1}$.\\
    Uncolor $\{x, f_k\}$.}
\End
\end{algorithmic}

\textit{Rotation lemma.} This operation maintains the properness of the edge coloring.
\begin{proof}
Consider some arbitrary $i < k$. We want to show that $c_{i+1}$ is a valid color for $\{x, f_i\}$ (i.e. that it maintains the properness of the edge coloring).

By definition, $c_{i+1}$ was free on $f_i$ before the fan was rotated.

Moreover, $\{x, f_i\}$ must be the only edge incident on $x$ colored $c_{i+1}$ after the rotation. Assume for a contradiction that there exists a vertex $v \neq f_i$ incident to $x$ such that $\{x, v\}$ is colored $c_{i+1}$ after the rotation.

\begin{itemize}[label=, leftmargin=*]
\item Case 1: $v \in F$. Then, $v = f_j$ for some $j \neq i$, and so $f_{j+1}$ was colored $c_{i+1}$ before the rotation. But this is not possible, since $f_{i+1}$ was colored $c_{i+1}$, and the edge coloring is proper.

\item Case 2: $v \notin F$. The rotation only affects edges contained in the fan, so $\{x, v\}$ must have been colored $c_{i+1}$ before the rotation. We also know that $f_{i+1}$ was colored $c_{i+1}$ before the rotation, and that $f_{i+1} \neq v$ since $v \notin F$. As earlier, this is not possible since the edge coloring is proper. 
\end{itemize}
\end{proof}

An \textit{alternating path} $P = \langle p_1, ... p_l \rangle$ on a vertex $x$ with colors $c, d$ is a nonempty sequence of distinct vertices such that $p_1 = x$, and for every $i < l$, the color of $\{p_i, p_{i+1}\}$ is either $c$ or $d$.

If this path is maximal (i.e. it cannot be extended on either side), we can invert the path without sacrificing the properness of the edge coloring. This may be useful to change the set of free colors on $x$, since if $c$ was free on $x$ prior to the inversion, then $d$ will be afterwards, and vice versa.

We describe the inversion procedure in pseudocode:

\begin{algorithmic}
\FunDef {invert} {$C[P]$}
    {For every $i < l$, flip the color of $\{p_i, p_{i+1}\}$. If it was colored $c$, recolor it with $d$; if it was colored $d$, recolor it with $c$.}
\End
\end{algorithmic}
\textit{Inversion lemma.} If $P$ is maximal, this operation maintains the validity of the edge coloring.
\vspace{-8pt}
\begin{proof}
Consider some arbitrary $i < l$, and suppose WLOG that $\{p_i, p_{i+1}\}$ was colored $c$. (The other case is symmetric, and follows by relabeling.) We want to show that $d$ is a valid color for $\{p_i, p_{i+1}\}$ post-inversion.

Since the edge coloring is proper, we know that:
\begin{enumerate}
\item{$\{p_{i-1}, p_{i}\}$ and $\{p_{i+1}, p_{i + 2}\}$ were both colored $d$ prior to the inversion.}
\item{No other edges incident to $p_i$ or $p_{i+1}$ were colored $d$.}
\end{enumerate}

After the inversion, $\{p_{i-1}, p_{i}\}$ and $\{p_{i+1}, p_{i + 2}\}$ are both colored $c$, leaving $d$ free on $p_i$ and $p_{i+1}$ as required.
\end{proof}

Another function that will be useful in describing the main algorithm is \textsc{recolor}, which sets the color of an edge within the edge coloring.

\begin{algorithmic}
\FunDef {recolor} {$C[\{u, v\} \mapsto a]$} {Returns the edge coloring given by $C$ but with the edge $\{u, v\}$ recolored with color $a$}
\End
\vspace{2pt}
\end{algorithmic}

We now describe a procedure \textsc{extend} $C \ U$, that starts with a partial $(\Delta + 1)$-edge coloring $C$, and a set $U$ which consists of the edges in $G$ that are uncolored in $C$. 

Choose an edge $\{x, y\}$ in $U$ to color. The main idea: if we can find a fan on $x$ whose first element is $y$ and whose last element shares some free color $u$ with $x$, we can simply rotate the fan to color $\{x, y\}$, and then color the last fan edge with $u$.

Our task is then to find such a fan. We start by constructing a maximal fan $F = \langle f_1, ..., f_k \rangle$ on $x$ whose first element is $y$. We are guaranteed that such a fan exists, since the singleton fan $\langle y \rangle$ is valid. We then choose a free color $c$ on $x$, and a free color $d$ on $f_k$. $c$ and $d$ must exist, since every vertex in $G$ can have at most $\Delta$ incident colors, and $C$ is a $(\Delta + 1)$-edge coloring.

If $d$ is free on $x$, we can rotate $F$, recolor $\{x, f_k\}$ with $d$, and then recursively color the remaining edges in $U - \{x, y\}$. 

However, if $d$ is incident on $x$, it must be made free. We construct an alternating path on $x$ with colors $c$ and $d$ by iteratively choosing edges colored $c$ and $d$ until the path cannot be extended any further. We then invert the path. Since $c$ was free on $x$ prior to the inversion, $d$ must be free on $x$ afterwards. 

We are not guaranteed that the path inversion preserves the properties of $F$, but it is possible to show that there exists a subfan of $F$ that exhibits the necessary properties -- i.e. that there is some $w \le k$ such that $\langle f_1, ..., f_w \rangle$ is a fan and $d$ is free on $f_w$. We will refer to this fact as the \textit{subfan existence lemma}, and justify it later.

We can then rotate the subfan, recolor $\{x, f_w\}$ with $d$, and recursively color the remaining edges.

Here is the algorithm in pseudocode:

\begin{algorithmic}[1]

\Fun {extend}{$U \ C$}
\If {$U = \emptyset$} $C$
\Else
    \State Choose $\{x, y\} \in U$. 
    \State Let $F = \langle f_1, ... f_k \rangle$ be a maximal fan on $x$ with $f_1 = y$. 
    \State Let $c$ be a free color on $x$, and $d$ be a free color on $f_k$. 
    \If {$d$ is an incident color on $x$}
        \State Let $P$ be a maximal alternating path on $x$ with colors $c, d$. 
        \State $C \defeq$ \textsc{invert} $C[P]$
    \EndIf
    \State Let $w \le k$ such that $F' = \langle f_1, ... f_w \rangle$ is a fan and $d$ is free on $f_w$. 
    \State $C' \defeq$ \textsc{rotate} $C[F']$ 
    \State $C' \defeq$ \textsc{recolor} $C'[\{x, f_w\} \mapsto d]$
    \State \textsc{extend} $(U -  \{x, y\})$ $C'$ 
\EndIf
\End
\end{algorithmic}

As promised, here is the proof of the subfan existence lemma:

\textit{Lemma.} The inversion of the alternating path guarantees the existence of a $w \le k$ such that $\langle f_1, ..., f_w \rangle$ is a fan and $d$ is free on $f_w$ (see: line 10 of \textsc{extend}).

\begin{proof}
In the case that $d$ is free on $x$, take $w = k$ and we're done. Otherwise:

Observe that $F$ must have exactly one edge colored $d$. It cannot have more than one, since each $f_i$ is a neighbor of $x$ and $C$ is a proper edge coloring. It also cannot have fewer than one, since otherwise we could append the neighbor of $x$ whose incident edge is colored $d$ to $F$, contradicting the maximality of $F$.

Let $\{x, f_i\}$ be this edge. 

\begin{enumerate}[label=\textit{Obs \arabic*.}, leftmargin=32pt]
    \item $i > 1$ since $f_1 = y$ and $\{x, y\}$ is uncolored.
    \item By definition of an alternating path, $x$ is an endpoint of $P$. Since $P$ is maximal, it must contain all of $x$'s neighbors that are colored $c$ or $d$. We can then conclude that $f_i \in P$. 
    \item $d$ is free on $f_{i-1}$; this is given by the definition of a fan since $\{x, f_i\}$ is colored $d$.
\end{enumerate}

\begin{itemize}[leftmargin=0pt, label=]
\item \textbf{Case 1:} $f_{i-1} \notin P$. 

    The path inversion only affects edges colored $c$ and $d$. $F$ does not have an edge colored $c$ (since $c$ is free on $x$). So, for all $j \neq i$, the free colors on $f_j$ and the color of $\{x, f_j\}$ are unchanged. Since $F$ was a fan, we can conclude that the fan color property still holds for all $f_j$.

    \vspace{4pt}
    Take $w = i - 1$. By Obs. 3, we are done.
\vspace{4pt}
\item \textbf{Case 2:} $f_{i-1} \in P$. Take $w = k$.
\vspace{4pt}

As earlier, we have that $f_j$ respects the fan color property for all $j \neq i$. By Obs. 2, $\{x, f_i\}$ is colored $c$ after the inversion. Since $d$ was free on $f_{i-1}$ (by Obs. 3), $f_{i-1}$ must be an endpoint of $P$, and so $c$ must be free on $f_{i-1}$ after the inversion.
\vspace{4pt}

Moreover, $f_k \notin P$. Suppose that it were; it would then have to be an endpoint since $d$ was free on $f_k$ prior to the inversion. However, this is impossible, because $x$ and $f_{i-1}$ are the endpoints of $P$.
\vspace{4pt}

We can therefore conclude that $d$ remains free on $f_k$, since it is unaffected by the inversion.  
\end{itemize}

\end{proof}

We can now finally describe the main algorithm \textsc{color} $G$, which takes in a graph and returns a complete proper edge coloring.

\begin{algorithmic}
\Fun {color} {$G$}
    \State Let $C$ be an empty $(\Delta + 1)$-edge coloring. 
    \State \textsc{extend} $E(G)$ $C$
\End
\end{algorithmic}

It is much easier to reason mathematically about purely functional programs than ones that contain side effects or mutable data. For this reason, \textsc{color} differs from the algorithm presented by Misra and Gries. 

We can now proceed to proving Vizing's theorem, which follows from:

\textit{Theorem.} \textsc{color} $G$ is a complete proper edge coloring of $G$ with at most $\Delta + 1$ colors.

\begin{proof}

By construction, \textsc{extend} only uses $\Delta + 1$ colors. We have already justified that each operation carried out in \textsc{extend} preserves the properness of its input edge coloring. We can thus apply an inductive argument to conclude that \textsc{color} $G$ is a proper edge coloring.

\textit{Obs.} At the start of each iteration of \textsc{extend}, $\{x, y\}$ must always be uncolored. To justify this, suppose that every edge in $U$ is uncolored in $C$. The only edges in $G$ whose colors are affected by \textsc{extend} are either in $P$, which contains no uncolored edges, or in $F$, whose only uncolored edge is $\{x, y\}$. Therefore, every edge in $U - \{x, y\}$ must be uncolored in $C'$. Applying this argument inductively yields the desired conclusion.

We may now proceed to proving completeness, using another inductive argument. It suffices to show that $C'$ contains one more colored edge than $C$.

Inverting the path does not affect the total number of edges colored, and neither does rotating the fan (provided that $\{x, y\}$ is uncolored, which we showed above). After line {\small (6)}, $\{x, f_k\}$ is uncolored, and line {\small (7)} colors it $d$, increasing the total number of edges colored by one as required.
\end{proof}

\section{Formalization: the library}

Many of the mathematical objects required to implement the algorithm have abstract definitions in Mathlib, Lean's open-source library of formalized mathematics. \cite{lean, mathlib} However, these definitions are not optimized for computational efficiency, so I chose to use Lean core data types such as lists and arrays instead. This allowed for a polynomial-time implementation of the algorithm, but required significant additional work --- in particular, I had to define:
\begin{itemize}
    \item A representation of each mathematical object that preserves the polynomial time complexity of the algorithm
    \item Implementations of key subroutines (recoloring an edge, rotating a fan, and inverting a path), and proofs that they preserve the invariants of the objects that they modify
    \item Formalizations of various notions (such as maximality) that are presented without explicit definition in the informal proof
\end{itemize}

The Lean code in this section is cleaned up, with some technical details omitted for readability. 

\subsection{Simple graphs}

Vertices are represented as natural numbers less than $n$, and edges are represented as pairs of vertices.
\begin{lstlisting}
variable (n : Nat) 
abbrev Vertex := Fin n 
abbrev Edge := Vertex n × Vertex n
\end{lstlisting}
The type \code{Fin n} represents natural numbers less than $n$.

A neighbor set is defined as a list of unique vertices whose length is less than $n$; the inequality is strict to disallow self loops. (\code{nodupAx} alone only implies that the length is less than or equal to $n$, so we impose the additional condition \code{sizeAx}.)

\begin{lstlisting}
structure Nbors (n : Nat) where
  val : List (Vertex n)
  sizeAx : val.length < n
  nodupAx : val.Nodup 
\end{lstlisting}

Here, \code{val.length} is shorthand for \code{List.length val}.

The predicate \code{Nodup} indicates that a vertex is not allowed to be present multiple times within a neighbor set -- i.e. multi-edges are not allowed.

Simple graphs are represented as adjacency lists, that map vertices to their neighborhoods. We also have the additional properties that every vertex is represented in the graph, that edges are bi-directional, and that there are no self loops. 
\begin{lstlisting}
structure Graph (n : Nat) where
  val : Array (Nbors n) 
  sizeAx : val.size = n 
  symmAx : ∀ u v : Vertex n, 
    u ∈ val[v] ↔ v ∈ val[u] 
  noSelfLoopsAx : ∀ u : Vertex n, u ∉ val[u]
\end{lstlisting}

Notably, edges are represented as tuples even though the graph is undirected. Ideally, the \code{Edge} type would have been unordered, but the type for unordered pairs in Mathlib (\code{sym2}) would have caused more messiness in theorem statements and required the use of many additional library lemmas over the course of the formalization. This choice did not result in much extra work -- I defined a few lemmas about symmetry and used them to rewrite statements wherever necessary. However, this could be a good candidate for using macros and refactoring.

The edge set of a graph is defined as a list of graph edges (i.e. a subset of $E(G)$), with a few additional properties:
\begin{itemize}
\item If $(u, v)$ is present in the edge set, $(v, u)$ must also be.
\item No multi-edges.
\end{itemize}

\subsection{Edge colorings}

\subsubsection{Definitions}

\code{Color} is defined as an option type: either a natural number less than $c$, or \code{none}. This is to allow for partial edge colorings.

\code{EdgeColoring c G} is a structure type parametrized by the graph being colored and the number of colors in the edge coloring. The data is represented as a modified adjacency matrix, where indices correspond to the vertices present in an edge pair and entries correspond to colors (either \code{none}, if the edge is uncolored, or \code{some c}, if the edge is colored with a color $c$).

We also have the following properties:
\begin{itemize}
\item The adjacency matrix is $n$ by $n$.
\item Every edge that is colored in the edge coloring is present in the graph.
\item No two incident edges have the same color (unless that color is none).
\item The matrix is symmetric -- \code{C[u][v] = C[v][u]}.
\end{itemize}

2D arrays have the nice computational property of constant-time access, but they can cause difficulty when writing statements and proofs, and when using automation to discharge goals within those proofs.

For example, in order to access the element at a given index of an array, Lean requires a proof that the index is in bounds. In a 2D array, this becomes particularly messy. Here is a function that returns the color of a given edge in an edge coloring:
\begin{lstlisting}
variable (C : EdgeColoring c G)

def color (e : Edge n) :=
  (C.val[e.1]'(by rw [C.sizeAx1]; exact e.1.isLt))[e.2]'
  (by rw [C.sizeAx2]; exact e.2.isLt)
\end{lstlisting}

I found that prioritizing computational efficiency foremost and then ease of using automation before ease of writing statements and proofs was the best strategy to write a relatively efficient implementation that did not require me to manually discharge as many difficult and technical goals. 

There are a number of other functions and lemmas about edge colorings in the Lean files, which I implemented for ease of proving later theorems. They are omitted here for brevity.

\subsubsection{Implementing the recolor operation}
The last important piece of the edge coloring library is the implementation of the \textsc{recolor} operation from the previous section.

We begin by formalizing the notion of validity. If a color $a$ is valid for edge $e$ (i.e. \code{edgeColorValid e a}), then either $a$ is \code{none}, or it is free on both vertices in $e$. 

We implement a function to set a cell in a 2D matrix to a given value. We also prove the following specifications:
\begin{itemize}
\item The size of the array remains invariant.
\item Every other value in the array remains invariant.
\item The value at the given cell is indeed the given value.
\item Every row in the array aside from the $i$th one remains invariant.
\item Setting two distinct cells in a 2D square matrix is a commutative operation.
\item We also include lemmas that allow us to reason about the count of a given value in the array before and after the operation.
\end{itemize}

Many list and array library functions have incomplete APIs; I tried to avoid using those functions wherever possible. For this reason, the library includes a specification of the \code{edgeColorValid} predicate that uses the function \code{Array.count}. This lemma states that $a$ is free on $\{u, v\}$ if and only if the count of $a$ is zero in both $C[u]$ and $C[v]$.

We can now define a function to recolor an edge, which sets both cells in the edge coloring that correspond with that edge. This operation comes with its own set of specifications to prove. Essentially, if we recolor edge $e = \{u, v\}$ with color $a$ in edge coloring $C$:
\begin{itemize}
\item The size and symmetry of $C$ are preserved.
\item $e$ is indeed colored with $a$, and the colors of the other edges remain invariant.
\item For any color $b$, the count of $b$ in the neighborhoods of $u$ and $v$ increases by one if $a = b$, decreases by one if $e$ was previously colored with $b$, and remains the same otherwise.
\end{itemize}

Lastly, to create an object of type \code{EdgeColoring}, we must prove that this operation maintains the properties listed in the definition of an edge coloring (size, representing the graph, properness, and symmetry). These proofs were long and technical since they required casework over indices and many invocations of symmetry lemmas; they are again omitted here.

\subsection{Fans}

\subsubsection{Definitions}
We define a fan $F$ on vertices $x$ and $y$ as a nonempty array of distinct vertices with the following additional properties:
\begin{itemize}
\item The first element of $F$ is $y$.
\item Every element of $F$ is in the neighborhood of $x$.
\item The fan color property holds as described in the previous section.
\end{itemize}

To implement the fan color property, we use the \code{Chain'} function in Lean, which indicates that a given predicate holds between adjacent elements of the list. This allows us to express an index-dependent property without actually using indices, which will save us a lot of proof obligations later.
\begin{lstlisting}
def fan_prop (x : Vertex n) (f₁ f₂ : Vertex n) :=
  color C (x, f₂) ∈ freeColorsOn C f₁

def colorAx (F : Array (Vertex n)) (x : Vertex n) :=
  F.toList.Chain' (fan_prop C x)
\end{lstlisting}

We also define the notion of a singleton fan on vertices $x$ and $y$, and prove that the fan properties hold under the assumption that the edge $\{x, y\}$ is present in the graph.

\subsubsection{Creating a maximal fan}
Given an uncolored edge $\{x, y\}$, the algorithm requires that we produce a maximal fan on $x$ whose first element is $y$. 

We start by defining a function \code{add} that, given a partial fan \code{F} and a list of neighbors \code{nbors} that are not currently present in the fan, adds as many elements to the fan as possible. 

\code{add} uses the \code{find?} function to search \code{nbors} for a vertex $z$ such that the color of $\{z, x\}$ is free on the last vertex in \code{F}. If such a $z$ exists, it is added to the back of \code{F} and erased from \code{nbors}. Otherwise, \code{F} is returned unchanged.

If \code{List.erase} is given a value that is not present in the list, it returns the list unchanged. So, in order to prove that \code{add} terminates, we must show that $z$ is always a member of \code{nbors} prior to its removal. By default, we do not have access to the definition of $z$ within the cases of the match statement. So, we use the function \code{List.attach} to bundle $z$ together with a proof that it is a member of \code{nbors}. This looks slightly less clean than adding a hypothesis to the match discriminant, but allows the simplifier and \code{split} to work more effectively on the match statement.

\begin{lstlisting}
def add (x y : Vertex n) (F : Array (Vertex n)) (nbors : List (Vertex n))
  (h : F ≠ #[]) :=
  match (List.attach nbors).find? (fun ⟨z, _⟩ ↦
    color C (x, z) ∈ freeColorsOn C (F.back (Array.size_pos_iff.mpr h))) with
  | some z => add x y (F.push z.val) (nbors.erase z.val) Array.push_ne_empty
  | none => F
  termination_by nbors.length
  decreasing_by
    rw [List.length_erase_of_mem z.prop]
    apply Nat.sub_one_lt_of_le
    apply List.length_pos_iff_exists_mem.mpr
    exact ⟨z.val, z.prop⟩
    rfl
\end{lstlisting}

We then state and prove lemmas which show that the fan properties are invariant under \code{add}. Since \code{add} is not structurally recursive, we induct over the function definition itself using the \code{fun_induction} tactic; this allows for simpler proofs despite the complex structure of the function. 

The maximality condition is formulated similarly to the function definition itself -- after calling \code{add} on a fan, every neighbor of $x$ not already present in the fan cannot be added to the back of the fan without violating the coloring invariant. Functional induction allows for a straightforward proof.

Finally, we define a wrapper function \code{maximalFan} that creates an object of type \code{Fan}. For the data, we call \code{add} on the singleton fan $\langle y \rangle$ and the neighbor set \code{nbhd x} $ - \{y\}$. For the properties, we invoke the lemmas we proved earlier about \code{add} along with the properties of the singleton fan.

\subsubsection{Implementing the rotate operation}

To rotate a fan $F = \langle f_1, ..., f_k \rangle$ on edge coloring $C$ with a given color $a$, we recolor $\{x, f_k\}$ with $a$, then recursively rotate the fan $\langle f_1, ..., f_{k-1} \rangle$ with the previous color of $\{x, f_k\}$. 

Since the formal definition of a fan is dependent on an edge coloring, we are required to prove that $\langle f_1, ..., f_{k-1} \rangle$ is still a fan on $C[\{x, f_k\} \mapsto a]$ before we can make the recursive call.

To accomplish this, we write a function \code{mkFan}, which takes in the fan $\langle f_1, ..., f_k \rangle$ on edge coloring $C$ (along with the color $a$ and a few additional hypotheses), and returns the fan $\langle f_1, ..., f_{k-1} \rangle$ on edge coloring $C[\{x, f_k\} \mapsto a]$.

\begin{lstlisting}
def mkFan (a : Color c)
  (hvalid : edgeColorValid C (x, last F) a) (hsize : F.val.size > 1):
  Fan (setEdgeColor C (x, last F) a (last_present F) hvalid) x y where
  ...
\end{lstlisting}

We can then use \code{mkFan} to implement the \code{rotateFan} operation as described above.

\begin{lstlisting}
def rotateFan (C : EdgeColoring c G) (F : Fan C x y) (a : Color c)
  (hvalid : edgeColorValid C (x, last F) a)
  : EdgeColoring c G :=
  let a' := color C (x, last F)
  let C' := setEdgeColor C (x, last F) a (last_present F) hvalid
  if h : F.val.size > 1 then
  let F' := mkFan F a hvalid h
  have hvalid' : edgeColorValid C' (x, last F') a' := by ...
  rotateFan C' F' a' hvalid'
  else C'
  termination_by ...
  decreasing_by ...
\end{lstlisting}

\subsection{Paths}

\subsubsection{Defining an alternating property}

Before we can state a definition for an alternating path, we must define the notion of a property that alternates over contiguous values in a list. \code{Chain'} is unfortunately not useful here, since we would need access to the indices of each element in order to describe the alternating aspect.

Instead, we recursively define a function \code{alternates}, which takes in a function $p$, two values $a, b$, and a list $vs$. \code{alternates p a b vs} indicates that the function $p$ alternates in value between $a$ and $b$ when evaluated over contiguous elements of the list. That is, if $vs = \langle v_1, v_2, v_3, ..., v_n \rangle$, $p \ v_1 \ v_2 = a$, $p \ v_2 \ v_3 = b$, and so on.

\begin{lstlisting}
def alternates {V C : Type*} (p : V → V → C) (a b : C) : List V → Prop
  | []               => True
  | _ :: []          => True
  | v₁ :: (v₂ :: vs) => p v₁ v₂ = a ∧ alternates p b a (v₂ :: vs)
\end{lstlisting}

We also define a function \code{next}, which returns the next value that $p$ should evaluate to, if an element were appended to the list that respects the alternating property. That is, for any element $w$, \code{alternates p a b (vs ++ [w])} if and only if \code{next a b vs} = $p \ v_n \ w$.

\begin{lstlisting}
def next {V C : Type*} (a b : C) : List V → C
  | []      => b
  | _ :: vs => next b a vs
\end{lstlisting}

We prove several properties of these functions, to more easily reason about them later. Here are a few examples, where $vs = \langle v_1, ..., v_n \rangle$ is an arbitrary list.

\begin{itemize}
\item \code{next a b vs} is always equal to either $a$ or $b$.
\item For some element $w$, if \code{alternates p a b vs} and $p \ v_n \ w$, \code{alternates p a b (vs ++ [w])}.
\item If \code{alternates p a b (x :: xs)}, then \code{alternates p b a xs}.
\end{itemize}

It is often useful to view a list in two different ways: as its inductive definition, and as a function that maps indices to values. This is possible in Lean, but it is not particularly natural to do both at the same time, and translating between representations often requires a lot of work. To reduce future messiness, we prove a few lemmas involving indices:

\begin{itemize}
\item For any list $L$, if \code{next a b L = a}, $p \ L[L.length - 2] \ L[L.length - 1] = b$, and vice versa (which we prove using mutual induction).
\item Let $0 < i < L.length - 1$. Then, $p \ L[i-1] \ L[i] = a$ and $p \ L[i] \ L[i+1] = b$, or vice versa.
\end{itemize}

There are also several library lemmas about \code{Chain'} that translate between these perspectives. These may be useful later, so we prove that if \code{alternates p a b L}, then \code{List.Chain' (fun x y ↦ p x y = a ∨ p x y = b) L}.

\subsubsection{Creating a maximal path}
Until now, we have defined each mathematical object as a structure type with data and invariants. This allows for better organized and more readable code, but has the unfortunate side effect of rendering many of Lean's automation tactics less effective (including induction, which will be necessary for many of the proofs in this section).

To work around this, we declare the data and invariants as variables now, prove some necessary lemmas, and then state \enquote{nice versions} of each lemma later.

\begin{lstlisting}
variable
  (P : List (Vertex n)) (a b : Color c) (hne : P ≠ [])
  (x : Vertex n)
  (hx : P[0]'(...) = x)
  (hfree : b ∈ freeColorsOn C x)
  (hnodup : P.Nodup)
  (ha : a.isSome)
  (hb : b.isSome)
  (hneq : a ≠ b)
\end{lstlisting}

Here, a path is represented as a list of vertices, such that the edge between incident elements is colored either $a$ or $b$. We require that $a$ and $b$ be actual colors (rather than \code{none}), and that the path contain no cycles. As in the informal definition, $x$ is the first element of the path. \code{hfree} is not a path invariant, but we can assume it here and supply it later when we create a maximal path in the algorithm.

The algorithm requires us to produce a maximal alternating path containing a vertex $x$ whose edges are colored with either $a$ or $b$, so our next task is to implement this.

For convenience, we define a predicate \code{alternatesColor} using \code{alternates}. We also define a function \code{nextVertex}, which finds a possible candidate for the next vertex to add onto an alternating path. The result must be connected to the last vertex in the path with an edge colored either $a$ or $b$. If such a vertex $z$ exists, \code{nextVertex} returns \code{some z}; otherwise, it returns \code{none}.

To prove maximality later, we must also show that any candidates for the next vertex cannot already be present in the path. This statement is never justified in the informal proof, but is complex and tedious (almost 75 lines long!) to formalize. Here is a sketch of the proof:

Let $P = \langle p_0 = x, ..., p_n\rangle.$ Let \code{some z} be the value returned by \code{nextVertex}. Assume for the sake of contradiction that $z$ is present in $P$. Then, there must exist some $i$ such that $z = p_i$. 
\begin{itemize}[label=, leftmargin=*]
\item \textbf{Case 1:} $i = 0$, so $z = p_0 = x$.
    Since $P$ is an alternating path, $\{p_n, x\}$ must be colored either $a$ or $b$. $b$ is free on $x$, so we can conclude $\{p_n, x\}$ is colored $a$, and $\{p_{n-1}, p_n\}$ must be colored $b$. 
    
    Since the edge coloring is proper, $x$ cannot have two incident edges colored $a$, and $p_n = p_1$. $P$ cannot contain duplicates, so $P = \langle x, p_1 \rangle$ and $p_{n-1} = x$. We reach a contradiction, since $\{x, p_1\}$ cannot be colored both $a$ and $b$.
\item \textbf{Case 2:} $0 < i < n$.
    Immediately, we know that $n > 1$.

    Since $P$ is an alternating path, $\{p_n, p_i\}$ must be colored either $a$ or $b$. Assume WLOG that $\{p_n, p_i\}$ is colored $a$. Since $p_i$ is an interior vertex, we have that $\{p_{i-1}, p_i\}$ is colored $a$, or that $\{p_i, p_{i+1}\}$ is colored $a$. In the former case, $n = i-1$, but this is impossible since $i < n$. So, suppose that $\{p_i, p_{i+1}\}$ is colored $a$. Then, $n = i+1$, and since the last path edge is colored $a$, the next one must be colored $b$. We reach a contradiction, since $\{p_n, p_i\}$ cannot be both colored $a$ and $b$.
\item \textbf{Case 3:} $i = n$, so $p_i = p_n$.
    $\{p_n, p_i\}$ is a self-loop, so it must be uncolored. But $\{p_n, p_i\}$ must be colored either $a$ or $b$, both of which cannot be \code{none}, so we reach a contradiction.
\end{itemize}

Armed with this lemma, we can define a function to extend an existing path until it is maximal. \code{extendPath} is similar in structure to \code{add} in the fan library (i.e. not structurally recursive, so we will have to manually prove termination). We first search for a candidate vertex $z$ to add to the path; if one does not exist, we return the path unchanged. Otherwise, we add $z$ to the path, and prove that each path invariant is still maintained before calling \code{extendPath} recursively on the resulting path.

\begin{lstlisting}
def extendPath {a b : Color c} {x : Vertex n}
  (P : List (Vertex n)) (hne : P ≠ [])
  (hx : P[0]'(by exact List.length_pos_iff.mpr hne) = x)
  (hfree : b ∈ freeColorsOn C x)
  (hnodup : P.Nodup)
  (ha : a.isSome) (hb : b.isSome) (hneq : a ≠ b)
  (hcolor : alternatesColor C P a b) : List (Vertex n) :=
  match Option.attach (nextVertex C P a b hne) with
  | none => P
  | some ⟨z, h⟩ =>
    have hnodup' : (P.concat z).Nodup := by ...
    have hcolor' : alternatesColor C (P.concat z) a b := by ...
    have hx' : (P.concat z)[0]'(by simp) = x := by ...
    extendPath (P.concat z) (by simp) hx' hfree hnodup' ha hb hneq hcolor'
  termination_by (n + 1) - P.length
  decreasing_by ...
\end{lstlisting}

Like earlier, we use functional induction to prove that the path invariants are preserved, and that this function produces a maximal path. The proofs are omitted here.

\subsubsection{Definitions}
We can now finally proceed to defining an alternating path as a structure type:

\begin{lstlisting}
structure Path (a b : Color c) (x : Vertex n) where
  val : List (Vertex n)
  nonemptyAx : val ≠ []
  firstElemAx : val[0] = x
  nodupAx : val.Nodup
  colorAx : alternatesColor C val a b
\end{lstlisting}

The strategy for creating a maximal path is similar to the version for fans: we call \code{extend} on a singleton path. We define a predicate for path maximality (that the next color in the path is free on its last vertex), and prove that the maximal path satisfies this condition.

\subsubsection{Reasoning about adjacent elements in a list}

Earlier in this section, I referenced a common challenge in proving theorems about lists -- that it is difficult to translate between viewing a list inductively, and viewing it as a function that maps indices to values. This was especially true when reasoning about adjacent pairs in a list, a notion that Mathlib does not already define. We will define it here.

First, a definition based on indices: if elements $u$ and $v$ are adjacent in a list $xs$, there must exist an index $i$ such that $u = xs[i]$ and $v = xs[i+1]$, or vice versa. 
\begin{lstlisting}
variable
  {α : Type*}
  {xs : List α}

def adjacent (u v : α) (xs : List α) :=
  (∃ i, ∃ (h : i < xs.length - 1), xs[i] = u ∧ xs[i+1] = v) ∨
  (∃ i, ∃ (h : i < xs.length - 1), xs[i] = v ∧ xs[i+1] = u)
\end{lstlisting}

Second, an inductive definition: given a list $xs$, \code{allAdjacentPairs xs} returns the list of each adjacent pair in $xs$.
\begin{lstlisting}
def allAdjacentPairs : List α → List (α × α)
  | [] => []
  | [_] => []
  | x₁ :: x₂ :: xs => (x₁, x₂) :: (x₂, x₁) :: (allAdjacentPairs (x₂ :: xs))
\end{lstlisting}

We prove a lemma that allows us to translate between these notions of adjacency. Functional induction is useful here, since the recursive cases do not mirror the inductive structure of the list.
\begin{lstlisting}
theorem mem_allAdjacentPairs_iff_adjacent (u v : α) (xs : List α) :
  (u, v) ∈ allAdjacentPairs xs ↔ adjacent u v xs := 
\end{lstlisting}

\subsubsection{Path edges}

By definition, $(u, v)$ is a path edge if $u$ and $v$ are adjacent in $P$. So, we define path edges as adjacent pairs in $P$:
\begin{lstlisting}
def pathEdges := allAdjacentPairs P.val
\end{lstlisting}

We prove a few more useful properties:
\begin{itemize}
\item Let $u, v$ be adjacent in $P$, and let $v$ be an interior vertex. There must exist some vertex $w$ adjacent to $v$, such that $(u, v)$ and $(u, w)$ are different colors.
\item If a vertex $v$ is a member of a maximal path, any of its incident edges colored $a$ or $b$ must be present in the path.
\end{itemize}

\subsubsection{Implementing the invert operation}
Our last task is to implement the \textsc{invert} operation described earlier. 

The simplest possible solution is to iteratively flip the color of each edge. Unfortunately, this will not work here, since we defined \code{setEdgeColor} to only allow edges to be recolored with valid colors. With a clever choice of hypotheses, we could recursively flip the tail of the path, then recolor the head. However, since the proof of correctness will require us to reason over multiple nested function calls, I chose a less elegant but simpler solution -- fully uncoloring the path, and then recoloring it in the opposite order. This intermediate step makes it easier to prove that the invariants are maintained.

We implement a function \code{uncolor}, which takes in a list of vertices and recursively sets the edge color of each adjacent pair to \code{none}. We also prove some auxiliary lemmas:
\begin{itemize}
\item Setting the color of a non-path edge does not affect the color invariant of the path.
\item If a color was free on a vertex prior to uncoloring the path, it must be free after.
\item After uncoloring the path, $a$ and $b$ must both be free on every path vertex.
\item Every non-path edge is unaffected by uncoloring the path.
\item After uncoloring the path, every edge in the path is colored \code{none}.
\end{itemize}

We can now proceed to defining \code{recolor}, which takes in an edge coloring $C$, colors $a, b$, and a list of vertices $L$, along with some hypotheses. If $L$ is an empty list or a singleton, we return the edge coloring unchanged. Otherwise, let $L = p_1 :: p_2 :: ps$. We call \code{recolor} recursively on edge coloring $C[(p_1, p_2) \mapsto b]$, colors $b, a$ (with order flipped), and list of vertices $p_2 :: ps$.

The one subtlety here is that we cannot carry through the hypothesis that both $a$ and $b$ are free on every vertex in $L$. This is true at the very start of the procedure, but once the first path edge is colored, $b$ is no longer free on $p_2$. We account for this asymmetry by a more careful choice of hypotheses: we are always guaranteed that the color we used last is free on the tail elements of $L$, and that the color we will use next is free on every element in $L$. This is all we need to show that each \code{setEdgeColor} operation is valid.

\begin{lstlisting}
def recolor (C : EdgeColoring c G) (a b : Color c) (ha : a.isSome) (hb : b.isSome)
  (hne : a ≠ b)
  (L : List (Vertex n)) (hnodup : L.Nodup)
  (h1 : List.Chain' (fun e₁ e₂ ↦ present G (e₁, e₂)) L)
  (h2 : ∀ v ∈ L.tail, a ∈ freeColorsOn C v)
  (h3 : ∀ v ∈ L, b ∈ freeColorsOn C v)
  : EdgeColoring c G :=
  match hl : L with
  | [] => C
  | [_] => C
  | p₁ :: p₂ :: ps =>
    have hpres : present G (p₁, p₂) := ...
    have hvalid : edgeColorValid C (p₁, p₂) b := ...
    have auxa : ∀ v ∈ p₂ :: ps,
      a ∈ freeColorsOn (setEdgeColor C (p₁, p₂) b hpres hvalid) v := ...
    have auxb : ∀ v ∈ (p₂ :: ps).tail,
      b ∈ freeColorsOn (setEdgeColor C (p₁, p₂) b hpres hvalid) v := ...
    recolor (setEdgeColor C (p₁, p₂) b hpres hvalid)
      b a hb ha (...) (p₂ :: ps) (...) (...) auxb auxa
\end{lstlisting}

We prove the following specifications: 
\begin{itemize}
\item If an edge is not present in the path, its color remains invariant.
\item The path alternates color between $b$ and $a$.
\end{itemize}

The main inversion function, \code{invert}, uncolors the path and then recolors it. 

\begin{lstlisting}
def invert {P : Path C a b x} (h : isMaximalPath P) : EdgeColoring c G :=
  recolor (uncolor C a b ha hb P.val P.nodupAx P.colorAx)
    a b ha hb hne P.val P.nodupAx (...) (...)
\end{lstlisting}

To prove its correctness, we start by defining the predicate \code{isInverted}, which takes in two edge colorings $C$ and $C'$, and a path $P$. \code{isInverted C C' P} indicates that for all edges:
\begin{itemize}
\item \code{isInverted_notmem C C' P}: if $e \notin P$, \code{color C e = color C' e}.
\item \code{isInverted_mem C C' P}: if $e \in P$:
\begin{itemize}
\item If \code{color C e = a}, then \code{color C' e = b}.
\item If \code{color C e = b}, then \code{color C' e = a}.
\end{itemize}
\end{itemize}

Lastly, we prove more specification lemmas about \code{invert}:
\begin{itemize}
\item The edge coloring returned by the \code{invert} function respects the \code{isInverted} predicate.
\item If an edge is not colored $a$ or $b$ prior to the inversion, it is not affected by the inversion.
\item If an edge is not colored $a$ or $b$ after the inversion, it was not affected by the inversion.
\item If a vertex is present in a maximal path and has an incident edge colored with a or b after the inversion, that edge must be present in the path.
\item If $a$ was free on a path vertex prior to the inversion, $b$ must be free after (and vice versa).
\end{itemize}
Many of these lemmas are variations of each other, for convenience in writing later proofs.

\section{Formalization: the algorithm}

The implementation of the algorithm is similar to the pseudocode in the informal proof: we define a subroutine \code{extendColoring}, which takes in a proper edge coloring and recursively extends it until it is complete. The main function, \code{mkEdgeColoring}, calls \code{extendColoring} on the empty edge coloring. 

A few notes:
\begin{itemize}
\item The proofs of each \code{have} statement are omitted for brevity.
\item To prove termination, we must show that every edge in $E$ is uncolored in $C$. This is proved by induction, similarly to its informal counterpart.
\item We will implement \code{findSubfan} and prove the subfan existence lemma in the next section.
\end{itemize}

Here are the implementations of both functions; they are not too different from the pseudocode in the informal proof!

\begin{lstlisting}
def extendColoring (E : EdgeSet G) (C : EdgeColoring (maxDegree G + 1) G)
  (h : ∀ e ∈ E.val, color C e = none) :
  EdgeColoring (maxDegree G + 1) G :=
match hE : E.val with
| [] => C
| (x, y) :: es =>
  have hterm : (remove E (x, y)).val.length < E.val.length 
  have h' : ∀ e ∈ (remove E (x, y)).val, color C e = none 
  
  let F := maximalFan C (...)
  have auxF : isMaximal F 
  
  let a := (freeColorsOn C (last F)).head (existsFreeColor C (by omega) (last F))
  let b := (freeColorsOn C x).head (existsFreeColor C (by omega) x)
  have ha : a.isSome
  have hb : b.isSome 
  have auxa : a ∈ freeColorsOn C (last F) 
  have auxb : b ∈ freeColorsOn C x 
  
  if heq : a = b then
    have aux : edgeColorValid C (x, last F) a 
    let C' := (rotateFan C F a aux)
    have h' : ∀ e ∈ (remove E (x, y)).val, color C' e = none 
    extendColoring (remove E (x, y)) (rotateFan C F a aux) h'
    
  else
    let P : Path C a b x := maximalPath C ha hb heq (by simp [b])
    have auxP : isMaximalPath P 
    let C' := invert ha hb heq auxb auxP
    have auxC' : isInverted C C' P 
    let F' := findSubfan ha hb auxb auxC' auxa (by simp_all) auxF auxP
    have aux : edgeColorValid C' (x, last F') a 
    let C'' := (rotateFan C' F' a aux)
    have h' : ∀ e ∈ (remove E (x, y)).val, color C'' e = none 
    extendColoring (remove E (x, y)) C'' h'
    
termination_by
  E.val.length

def mkEdgeColoring (G : Graph n) : EdgeColoring (maxDegree G + 1) G :=
  extendColoring (toEdgeSet G) empty (...)
\end{lstlisting}

\subsection{Finding the subfan}

The inversion of the alternating path guarantees the existence of a subfan of $F$, such that $a$ is free on its last element.

The informal proof of the subfan existence lemma is subtle, so to complete the formalization we must prove the following statements:
\begin{itemize}
\item If a vertex is not a path endpoint, its free colors are unaffected by the path inversion.
\item If a color not equal to $a$ or $b$ was free on a vertex prior to the path inversion, it remains free after the inversion.
\item If the path is empty, the color property of a fan is preserved. (This is trivially true since the inversion does nothing, but requires a little bit of work to formalize.)
\item Suppose there exists a fan edge $e$ with color $a$. Then, every edge in the fan not equal to $e$ is unchanged by the path inversion. 
\item The maximal path containing $x$ is a singleton if there does not exist a fan edge colored $a$.
\end{itemize}

The statement and proof of the lemma is omitted here, since it follows the same structure as its informal counterpart but required extensive casework and arithmetic. Control structures (especially tactics that allowed me to work on multiple goals at once), and specialized automation (such as \code{omega} and \code{grind}) were particularly helpful.

Now that we have proved the existence of the subfan, we can proceed to finding it. This task is straightforward -- if there is an index $i$ such that $\{x, f_i\}$ is colored $a$ and $f_{i-1} \notin P$, we return the first $i-1$ elements of $F$. Otherwise, we return $F$ unchanged. In Lean:
\begin{lstlisting}
def mkSubfan (F : Fan C x y) (P : Path C a b x) : Array (Vertex n) :=
  match Array.findFinIdx? (fun u ↦ color C (x, u) = a) F.val with
  | some i => if F.val[i.val - 1] ∈ P.val then F.val else Array.extract F.val 0 i
  | none => F.val
\end{lstlisting}

We can now write a \enquote{nice} version of this function that returns a \code{Fan} object (i.e. that contains a proof of each fan property). The last property left to prove is that $a$ is always free on the last vertex in the subfan. The proofs are omitted here for brevity.

\subsection{The proof of correctness}
In the informal proof, the correctness argument had two main components: properness and completeness. The first part is already done -- \code{mkEdgeColoring} produces an object of type \code{EdgeColoring}, which is proper by definition! All that is left is proving completeness.

We use a familiar strategy here -- first, we prove that \code{extendColoring} preserves properness (i.e. that every edge we have removed from the edge set is colored in the graph).

\begin{lstlisting}
theorem extendColoring_proper (E : EdgeSet G) (C : EdgeColoring (maxDegree G + 1) G)
  (he1 : ∀ e ∈ E.val, color C e = none)
  (e : Edge n) (h1 : e ∈ (toEdgeSet G).val)
  (h2 : e ∉ E.val → (color C e).isSome) :
  (color (extendColoring E C he1) e).isSome :=
\end{lstlisting}

The proof is similar to its informal counterpart (but requires the rotation lemmas we proved earlier), so we omit it here.

We can now state the last proof of correctness: that \code{mkEdgeColoring} creates a proper edge coloring.

\begin{lstlisting}
theorem mkEdgeColoring_proper :
  ∀ e ∈ (toEdgeSet G).val, (color (mkEdgeColoring G) e).isSome := 
\end{lstlisting}

The proof follows almost immediately from the previous lemma, and we are done.

\section{Conclusion}

In this work, I wrote a polynomial-time verified implementation of the Misra and Gries edge coloring algorithm in the proof assistant Lean 4. The algorithm was initially developed to simplify the proof of a complex theorem, so having a computer-verified proof of correctness is exciting! 

Using an interactive theorem prover was essential -- many of the proofs required subtle arguments that would be difficult for fully automated tools to synthesize. However, this choice presented its own set of challenges:
\begin{itemize}
\item A common strategy in mathematical proof is to translate between different representations of the same object. For example, within a single proof, I often needed to view a list both inductively and as a function mapping indices to values. Such translations are more challenging in formalized proofs, but I addressed this by writing dedicated lemmas to carry them out wherever they were needed.
\item Formalizing a mathematical algorithm introduces competing objectives that have to be balanced during implementation: representations that are easy to reason about mathematically are not always the most computationally efficient, and vice versa. I chose to prioritize computational efficiency, since it is an inherent property that cannot be improved without changing the underlying representations. To manage the messy and complex proofs that resulted, I relied on a variety of strategies, particularly automation wherever possible.
\item These considerations raise the question of whether it would be easier to use entirely separate representations for computation and for proof, and then show that they are equivalent. I suspect the proofs would be simpler -- especially if I used the definitions in Mathlib -- though this would require a significant amount of additional translation back and forth.
\end{itemize}
There are many possible directions for future work. The library could be helpful to implement more verified graph-theoretic algorithms with nice computational properties. In addition, there are many places where macros, custom automation, or different choices of representation could have allowed for shorter and more elegant proofs. As more work like this is completed, I am excited to see best practices emerge for representing and reasoning about mathematical objects.

\printbibliography

@article{misragries,
title = {A constructive proof of Vizing's theorem},
journal = {Information Processing Letters},
volume = {41},
number = {3},
pages = {131-133},
year = {1992},
issn = {0020-0190},
doi = {https://doi.org/10.1016/0020-0190(92)90041-S},
url = {https://www.sciencedirect.com/science/article/pii/002001909290041S},
author = {J. Misra and David Gries}
}

@book{bollobas,
  title={Graph theory: an introductory course},
  author={Bollob{\'a}s, B{\'e}la},
  volume={63},
  year={2012},
  publisher={Springer Science \& Business Media}
}

@InProceedings{raodijkstra,
author="Rao, Josyula R.
and Dijkstra, Edsger W.",
editor="Broy, Manfred",
title="Designing the proof of Vizing's Theorem",
booktitle="Programming and Mathematical Method",
year="1992",
publisher="Springer Berlin Heidelberg",
address="Berlin, Heidelberg",
pages="17--25",
isbn="978-3-642-77572-7"
}

@article{npcomplete,
author = {Holyer, Ian},
title = {The NP-Completeness of Edge-Coloring},
journal = {SIAM Journal on Computing},
volume = {10},
number = {4},
pages = {718-720},
year = {1981},
doi = {10.1137/0210055},

URL = { 
    
        https://doi.org/10.1137/0210055
    
    

},
eprint = { 
    
        https://doi.org/10.1137/0210055
    
    

}
}

@misc{breitner, title={Proving Vizing’s Theorem with Rodin}, url={https://www.joachim-breitner.de/}, author={Breitner, Joachim}, year={2011}, month={Mar}}

@article{vizing,
title = {Critical graphs with given chromatic class},
journal = {Metody Diskret. Analiz.},
volume = {5},
pages = {9–17},
year = {1965},
author = {Vizing, V. G.}
}

@article{nipkow,
author={Nipkow, Tobias},
title={Gale-Shapley Verified},
journal={Journal of Automated Reasoning},
year={2024},
month={Jun},
day={14},
volume={68},
number={2},
pages={12},
issn={1573-0670},
doi={10.1007/s10817-024-09700-x},
url={https://doi.org/10.1007/s10817-024-09700-x}
}

@article{tarjan,
  author       = {Ran Chen and
                  Cyril Cohen and
                  Jean{-}Jacques L{\'{e}}vy and
                  Stephan Merz and
                  Laurent Th{\'{e}}ry},
  title        = {Formal Proofs of Tarjan's Algorithm in Why3, Coq, and Isabelle},
  journal      = {CoRR},
  volume       = {abs/1810.11979},
  year         = {2018},
  url          = {http://arxiv.org/abs/1810.11979},
  eprinttype    = {arXiv},
  eprint       = {1810.11979},
  bibsource    = {dblp computer science bibliography, https://dblp.org}
}

@InProceedings{lean,
author="Moura, Leonardo de
and Ullrich, Sebastian",
editor="Platzer, Andr{\'e}
and Sutcliffe, Geoff",
title="The Lean 4 Theorem Prover and Programming Language",
booktitle="Automated Deduction -- CADE 28",
year="2021",
publisher="Springer International Publishing",
address="Cham",
pages="625--635",
isbn="978-3-030-79876-5"
}

@inproceedings{mathlib,
author = {The mathlib Community},
title = {The lean mathematical library},
year = {2020},
isbn = {9781450370974},
publisher = {Association for Computing Machinery},
address = {New York, NY, USA},
url = {https://doi.org/10.1145/3372885.3373824},
doi = {10.1145/3372885.3373824},
booktitle = {Proceedings of the 9th ACM SIGPLAN International Conference on Certified Programs and Proofs},
pages = {367–381},
numpages = {15},
location = {New Orleans, LA, USA},
series = {CPP 2020}
}

@software{mycode,
author = {Bhoja, Arohee},
title = {{A verified implementation of the Misra and Gries edge coloring algorithm}},
url = {https://github.com/aroheebhoja/vizing}
}
\end{document}